# Quantum friction

Roumen Tsekov
Department of Physical Chemistry, University of Sofia, 1164 Sofia, Bulgaria

The Brownian motion of a light quantum particle in a heavy classical gas is theoretically described and a new expression for the friction coefficient is obtained for arbitrary temperature. At zero temperature it equals to the de Broglie momentum of the mean free path divided by the mean free path. Alternatively, the corresponding mobility of the quantum particle in the classical gas is equal to the square of the mean free path divided by the Planck constant. The Brownian motion of a quantum particle in a quantum environment is also discussed.

Quantum friction is a phenomenon in which the energy dissipation appears only due to quantum effects. Hence, it has no analog in classical physics. Traditionally, the quantum friction is associated by contact free interactions due to the Casimir-Polder forces[1,2,3] but there is extensive literature devoted also to quantum friction in van der Waals and other systems.[4,5,6,7,8] Another example of quantum friction appears in the quantum Brownian motion at zero temperature,[9,10] which is important to the third law of thermodynamics.[11] An analogy between the quantum dissipative Brownian motion and the Casimir effect is also explored.[12] Conventionally, the quantum friction in the quantum Brownian motion originates from the quantum nature of the quantum bath, surrounding the Brownian particle. In the present paper an alternative aspect of quantum friction is discussed, which appears due to collisions of a quantum particle spreading in a classical gas environment at zero temperature.

The aim of the present study is to describe the Brownian motion of a light particle with mass $m$, moving among heavy gas particles with mass $M \gg m$. The motion of the target particle is stochastic, due to the random nature of interaction with the gas particles. The Brownian particle dynamics is rigorously described via the Langevin equation[13]

$$m\ddot{r} + b\dot{r} = f \tag{1}$$

where $r(t)$ is the Brownian particle coordinate and $f(t)$ is the random Langevin force. The second term in this stochastic Newtonian equation describes the friction force with friction coefficient $b$. Multiplying Eq. (1) by the $x$ vector component of the Brownian particle coordinate and taking the average value of the scalar products yields

$$m<x\ddot{x}> + b<x\dot{x}> = 0 \tag{2}$$

The contribution of the Langevin force vanishes here since $f$ is zero centered and not correlated with the Brownian particle position,[13] i.e. $<rf>=<r><f>=0$. Introducing the dispersion $\sigma_x^2 \equiv <x^2>$ of the $x$ scalar component of the Brownian particle coordinate, Eq. (2) changes after rearrangements to

$$m\partial_t^2 \sigma_x^2 + b\partial_t \sigma_x^2 = 2m\sigma_v^2 \qquad (3)$$

where $\sigma_v^2 \equiv <\dot{x}^2>$ is the $x$ component of the Brownian particle velocity dispersion. If the observation time is larger than the velocity relaxation time $\tau \equiv m/b$ the first inertial term can be neglected and Eq. (3) reduces to the virial theorem of the Brownian motion[14]

$$b\partial_t \sigma_x^2 = 2m\sigma_v^2 \qquad (4)$$

At equilibrium in the momentum space the velocity dispersion $\sigma_v^2 = k_B T/m$ is a constant, proportional to temperature. Integrating now Eq. (4) yields the classical Einstein law $\sigma_x^2 = 2Dt$ of the Brownian motion with diffusion constant given by the Einstein formula $D \equiv k_B T/b$.

The consideration above is valid for both gases and liquids. In the latter case the friction coefficient is given by the Stokes law $b = 6\pi\eta R$ and, thus, one arrives to the Stokes-Einstein relation $D = k_B T/6\pi\eta R$. In ideal gases, however, the interaction between the particles is due to binary elastic collisions only and all transport properties are described by the mean free path $\lambda \equiv 1/\sigma n$. Here $\sigma$ is the collision cross-sectional area, $n$ is the gas particle density, and the usual factor $\sqrt{2}$ is missing there, since the velocity of the light Brownian particle is much higher than that of the heavier gas particles. A textbook problem to solve now is what the friction and diffusion coefficients of the light particle are. Interpreting the relaxation time $\tau$ as the collision time one expects the following relation to hold, $\sigma_x^2(\tau) \approx \lambda^2$, since the mean free path is the definite length scale of collisions. Employing here the Einstein law results in an expression for the friction constant $b \approx \sqrt{mk_B T}/\lambda$. Clearly, the corresponding collision time $\tau \approx \lambda/\sqrt{k_B T/m}$ equals to the time needed by the target particle to travel between two consecutive collisions ballistically with the thermal velocity. For the diffusion coefficient one obtains further the classical expression $D \approx \lambda\sqrt{k_B T/m}$ from the gas kinetic theory.[15] As is seen, both the friction and diffusion coefficients vanish at zero temperature, because at $T=0$ the classical systems are in total rest. However, at low temperatures quantum effects become important and we are going to explore now how the light quantum particle diffuses in the heavy classical gas. The opposite problem of a heavy quantum Brownian particle moving in a light ideal gas with $M \ll m$ is al-ready described via the linear Boltzmann equation.[16] The obtained thermo-quantum diffusion constant $D \approx D_{cl} + (\hbar/4m)^2/D_{cl}$ is semiclassical since it diverges at $T = 0$.

At the considered short $\tau$ the collisions are very frequent and the target quantum particle is continuously measured by the gas particles. In this case the minimal Heisenberg relation $\sigma_v \sigma_x = \hbar/2m$ holds at any time. Expressing from here the Brownian particle velocity dispersion and introducing it into the virial theorem (4) lead, after integration over time, to[17]

$$\sigma_x^2 = \hbar\sqrt{t/mb} \tag{5}$$

This sub-diffusive law[18] shows that the quantum diffusion at zero temperature does not obey the Einstein law. Nevertheless, since the relaxation mechanism involves collisions only the collisional hypothesis $\sigma_x^2(\tau) \approx \lambda^2$ holds in the quantum case as well. Introducing it in Eq. (5) results in determination of the residual friction coefficient of the quantum particle in a classical gas

$$b \approx \hbar/\lambda^2 = \hbar(\sigma n)^2 \tag{6}$$

Hence, the mobility of a quantum particle in a classical gas at zero temperature is $\lambda^2/\hbar$. The corresponding collision time $\tau \equiv m/b \approx m\lambda^2/\hbar$ has a clear meaning; it is the ratio between the mean free path $\lambda$ and its corresponding de Broglie velocity $\hbar/\lambda m$. Note that since $\hbar/2m$ has dimension of a diffusion coefficient, the quantum relation $\lambda^2 \approx \hbar\tau/m$ hints a diffusive process in contrast to the classical ballistic expression $\lambda \approx \tau\sqrt{k_B T/m}$. For hydrogen atoms in solids $\lambda$ is commensurable with the lattice constant and the collision time $\tau$ is of the order of picoseconds according to Eq. (6).

Using the explicit expression (6) the sub-diffusive law (5) acquires an alternative form

$$\sigma_x^2 \approx \lambda\sqrt{\hbar t/m} \tag{7}$$

This equation has a bizarre meaning. Initially the particle runs 'mentally' a sequence of $N$ uncorrelated collisions, which corresponds to $\sigma_x^2 \approx \lambda^2 N$ according to stochastic laws. If the particle is classical, it travels later ballistically on this trajectory. Therefore, its travel time equals to $t = N\lambda/\sqrt{k_B T/m}$, which results in the Einstein law $\sigma_x^2 \approx (\lambda\sqrt{k_B T/m})t$. In contrast, a quantum particle does a normal diffusion on the 'mental' trajectory with the universal quantum diffusion coefficient $\hbar/2m$.[19] Hence, its travel time is $t = (N\lambda)^2/(\hbar/m)$, which yields straightforwardly Eq. (7). Therefore, at zero temperature the quantum particle undergoes a dual diffusive process. It is interesting to calculate the classical diffusion coefficient for the quantum particle

$$D(t) \equiv \sigma_x^2 / 2t \approx \lambda \sqrt{\hbar / 4mt} \qquad (8)$$

which is time-dependent. The structure of Eq. (8) resembles the classical diffusion coefficient $D \approx \lambda \sqrt{k_B T / m}$ with an effective temperature $T(t) \equiv \hbar / 4k_B t$ given by the minimal Heisenberg time-energy uncertainty relation. Using this time-dependent quantum temperature, the length of the 'mental' path equals to the thermal de Broglie wavelength $N\lambda = \hbar / 2\sqrt{mk_B T(t)}$, while the time-dependent velocity dispersion of the quantum particle $\sigma_v^2 = (\hbar / 2m\sigma_x)^2 \approx Nk_B T(t)/m$ is $N$ times larger than the classical analog.

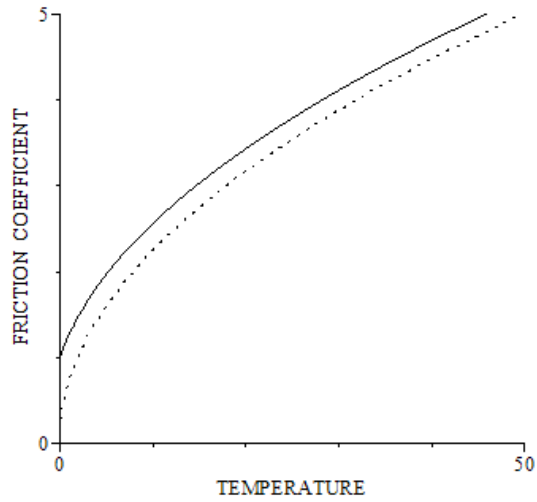

**Fig. 1** Dependence of the dimensionless friction coefficient $b\lambda^2 / \hbar$ on the dimensionless temperature $T / T_\lambda$; the solid line is according to Eq. (10) and the dotted line is for the classical one

It is also possible to obtain the friction coefficient of the quantum particle at arbitrary temperature. In this case the quantum Brownian motion is described via[17]

$$\sigma_x^2 - \lambda_T^2 \ln(1 + \sigma_x^2 / \lambda_T^2) = 2Dt \qquad (9)$$

where $\lambda_T \equiv \hbar / 2\sqrt{mk_B T}$ is the standard thermal de Broglie wavelength. Employing again the collisional hypothesis $\sigma_x^2(\tau) \approx \lambda^2$ results in

$$b \approx \sqrt{mk_B T / [\lambda^2 - \lambda_T^2 \ln(1 + \lambda^2 / \lambda_T^2)]} \approx (\hbar / \lambda^2) / \sqrt{(T_\lambda / T)[1 - (T_\lambda / T)\ln(1 + T / T_\lambda)]} \qquad (10)$$

This expression reduces to Eq. (6) at zero temperature, while at high temperature it provides the classical limit. As is shown in Fig. 1, the friction coefficient increases monotonously with temperature increase and the quantum/classical transition is marked by a characteristic temperature $T_\lambda \equiv \hbar^2 / 4m\lambda^2 k_B \approx T(\tau)$. The latter is of the order of 1 K for hydrogen atoms in solids.

Using Eq. (10) one can calculate the Einstein diffusion constant from Eq. (9). Expanding the result in a semiclassical series on the Planck constant leads to

$$D \equiv k_B T / b \approx D_{cl} - (\hbar / 4m)^2 / D_{cl} + o(\hbar^4) \tag{11}$$

As is seen, the structure of the quantum correction is similar to that derived by Vacchini and Hornberger[16] but the sign is opposite. The quantum effect increases the friction coefficient in Eq. (10) and thus it decreases the Einstein diffusion constant. However, the quantum effect intensifies the Brownian motion driving force as well and the result of Vacchini and Hornberger corresponds to the semiclassical limit of diffusion constant $k_B(T+T_\lambda)/b \approx \lambda\sqrt{k_B(T+T_\lambda)/m}$. Thus, in total the quantum effect increases the particle diffusivity. The temperature dependence of the friction constant leads to a slight temperature dependence of the pre-exponential factor in the Arrhenius law.[20]

The description above is valid for a quantum Brownian particle in a classical environment. The expression (9) is derived by Eq. (4) accomplished by the Maxwell-Heisenberg velocity dispersion $\sigma_v^2 = k_B T / m + (\hbar / 2m\sigma_x)^2$. The first Maxwell term vanishes at zero temperature but, if the environment is quantum, one should replace it by the effective quantum temperature to obtain $\sigma_v^2 = k_B T(t) / m + (\hbar / 2m\sigma_x)^2$. Introducing now this expression in Eq. (4) the virial theorem of the Brownian motion at zero temperature changes to

$$b\partial_t \sigma_x^2 = \hbar^2 / 2m\sigma_x^2 + \hbar / 2t \tag{12}$$

If the environment is classical, the last term omits and the integration of Eq. (12) leads to Eq. (5). In the case of a classical particle with large mass moving in a quantum environment the first quantum term is negligible. The integration of the resulting equation yields the known expression[21] $\sigma_x^2 = (\hbar / b)[\ln(\sqrt{bt/m}) + 1]$, which is positive since Eq. (12) is valid for $t \geq \tau$. The numerical solution of Eq. (12) with initial condition $\sigma_x^2(\tau) = \hbar / b$ is plotted in Fig. 2. As is seen, the exact result lays between the quantum-classical dispersion $\sigma_x^2 = \hbar\sqrt{t/mb}$ and quantum-quantum superposition $\sigma_x^2 = \hbar\sqrt{t/mb} + (\hbar / b)[\ln(\sqrt{bt/m}) + 1]$. Thus, a good approximation of the solution of Eq. (12) is $\sigma_x^2 = (\hbar / b)\{\sqrt{bt/m} + [\ln(\sqrt{bt/m}) + 1]/3\}$. Since this equation does not affect the friction coefficient from Eq. (6), in an ideal quantum gas it specifies further to

$$\sigma_x^2 \approx \lambda\sqrt{\hbar t/m} + \lambda^2[\ln(\sqrt{\hbar t/m}/\lambda) + 1]/3 \tag{13}$$

Note that the fist quantum-classical term depends linearly on the mean free path, while the last classical-quantum term has a more complicated dependence on $\lambda$. The dependence of the two terms on the particle mass is also different.

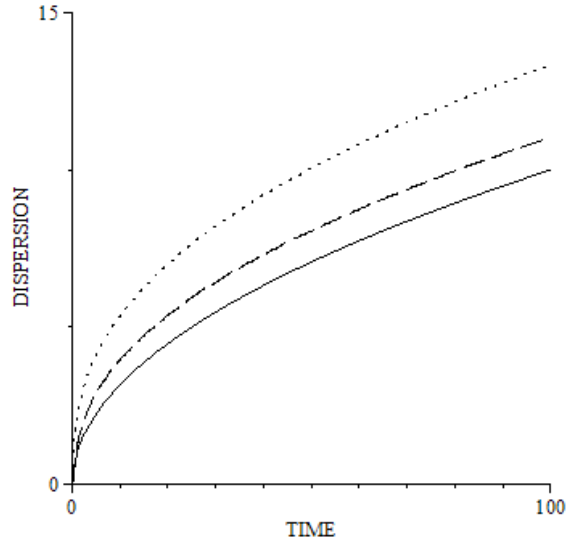

**Fig. 2** The dependence of the dimensionless dispersion $\sigma_x^2 b/\hbar$ on the dimensionless time $t/\tau$; the solid line is according to Eq. (5), the dashed line is according to Eq. (12) and the dotted line is for the quantum-quantum superposition $\sqrt{t/\tau}+\ln(\sqrt{t/\tau})+1$

Another interesting expression is derived in the following way. Because the leading term on the right-hand side of Eq. (12) is the first one, one can express the time from Eq. (7). Introducing the resultant $t = m\sigma_x^4/\hbar\lambda^2$ in the last term of Eq. (12) yields

$$\partial_t \sigma_x^2 = \hbar\lambda^2/2m\sigma_x^2 + \hbar\lambda^4/2m\sigma_x^4 \qquad (14)$$

The integration of this equation leads to

$$\sigma_x^4/\lambda^2 - 2\sigma_x^2 + 2\lambda^2 \ln(1+\sigma_x^2/\lambda^2) = \hbar t/m \qquad (15)$$

which is a combination between Eq. (7) and Eq. (9) with $\lambda_T = \lambda$. At large time Eq. (15) reduces to Eq. (7), while at short time it predicts a new quantum-quantum sub-diffusive law

$$\sigma_x^2 = \lambda\sqrt[3]{3\lambda\hbar t/2m} = \hbar\sqrt[3]{3t/2mb^2} \qquad (16)$$

Considering this equation and Eq. (5) one unveils a general fractional law $\sigma_x^2 \approx (\hbar/b)(bt/m)^{2\alpha}$, where $0 \leq \alpha \leq 1$ is a parameter reflecting different types of quantum diffusion. For gases according to Eq. (6) the root-mean-square displacement equals to $\sigma_x \approx \lambda(\hbar t/m\lambda^2)^\alpha$.

As is well known, quantum systems are not in rest at zero temperature, which manifests itself by the non-zero vacuum energy. Hence, according to the present study the interaction of an electron, for instance, with a frozen classical surrounding will result in appearance of a quantum friction. This effect should be also present even in atoms and molecules, where the contemporary quantum mechanics does not consider, surprisingly, any collisions between the building particles. Since the electrons are negatively charged, the electron-electron collisions are probably very rare events but they could be important for electron transitions in atoms and molecules.[22] The collisions between an electron and relatively large positive nuclei seem, however, inevitable unless it is restricted by relativistic effects.

## Appendix

The aim of the Appendix is to describe in more details the Brownian motion of a quantum oscillator fluctuating in a quantum bath at zero temperature. The oscillator coordinate $x$ obeys the following Bohm-Langevin equation[23]

$$m\ddot{x} + b\dot{x} + m\omega_0^2 x - \hbar^2 x/4m\sigma_x^4 = f \qquad (17)$$

where $\omega_0$ is the oscillator own frequency and $f$ is the quantum Langevin stochastic force. The last term on the left hand side represents the Bohm quantum force. For large time $t \gg m/b$ one can neglect the inertial term as compared to the friction and such simplified Eq. (17) can be easily integrated to obtain the integral solution

$$x = \frac{1}{b}\int_0^t \exp[\frac{1}{b}\int_{t_1}^t (\frac{\hbar^2}{4m\sigma_x^4} - m\omega_0^2)ds_1] f(t_1) dt_1 \qquad (18)$$

It is straightforward now to calculate the oscillator dispersion in the form

$$\sigma_x^2 = \frac{1}{b^2}\int_0^t\int_0^t \exp[\frac{1}{b}\int_{t_1}^t (\frac{\hbar^2}{4m\sigma_x^4} - m\omega_0^2)ds_1 + \frac{1}{b}\int_{t_2}^t (\frac{\hbar^2}{4m\sigma_x^4} - m\omega_0^2)ds_2] C_{ff}(t_2 - t_1) dt_1 dt_2 \qquad (19)$$

Using the definition $k_B T(t) \equiv \hbar/4t$ of the effective quantum temperature, the autocorrelation function of the quantum Langevin force acquires the form

$$C_{ff}(t) \equiv < f(t)f(0) > = 2bk_B T(t)\delta(t) = b\hbar\delta(t)/2t \tag{20}$$

with $t \geq 0$. Introducing this expression in Eq. (19) and performing the integrations yields an integral equation for the oscillator dispersion

$$\sigma_x^2 = \frac{\hbar}{b}\{\exp[\frac{2}{b}\int_0^t (\frac{\hbar^2}{4m\sigma_x^4} - m\omega_0^2)ds] - 1\} \tag{21}$$

The quadratic $\hbar^2$-term originates from the Bohm quantum potential and accounts for the quantum nature of the oscillator, while the linear $\hbar$-term follows from the quantum nature of the bath. By simple differentiation in time Eq. (21) transforms into differential form

$$b\partial_t \sigma_x^2 = (\hbar^2/2m\sigma_x^4 - 2m\omega_0^2)(\sigma_x^2 + \hbar/b) \tag{22}$$

As is seen, the equilibrium dispersion of the quantum harmonic oscillator at zero temperature is given by the traditional expression $\sigma_x^2 = \hbar/2m\omega_0$. In the case of a free particle ($\omega_0 = 0$) Eq. (22) reduces to Eq. (14). It can be integrated to obtain an alternative form of Eq. (15)

$$(b\sigma_x^2/\hbar)^2 - 2(b\sigma_x^2/\hbar) + 2\ln(1 + b\sigma_x^2/\hbar) = bt/m \tag{23}$$

At large time the particle position dimensionless dispersion tends to the result $b\sigma_x^2/\hbar = \sqrt{bt/m}$ from Eq. (5), while the expression $b\sigma_x^2/\hbar = \sqrt[3]{3bt/2m}$ from Eq. (16) holds at short time.